\documentclass[%
 reprint,
 amsmath,amssymb,
 aps,
]{revtex4-2}
\DeclareRobustCommand{\VAN}[3]{#2}
\let\VANthebibliography\thebibliography
\def\thebibliography{\DeclareRobustCommand{\VAN}[3]{##3}\VANthebibliography}

\usepackage{graphicx}
\usepackage{dcolumn}
\usepackage{bm}

\usepackage{color,float}

\begin{document}


\title{Locality in collider tests of Quantum Mechanics with top quark pairs }

\author{Regina Demina}
\author{Gabriel Landi}
\affiliation{Department of Physics and Astronomy, University of Rochester, 206 Bausch and Lomb Hall, Rochester, NY}
\email{regina@pas.rochester.edu}

\date{\today}

\begin{abstract}
Tests of quantum properties of fundamental particles in high energy colliders are starting to appear. However, such experiments may suffer from the locality loophole. We argue for criteria that take into account the space-like separation between measurements for the case of spin correlations in top quark pairs produced at the LHC. We derive bounds considering three different definitions of what constitutes the quantum measurement - the decay of top quarks, the decay of $W$ bosons, and the stable decay products contacting a macroscopic device.
\end{abstract}

\maketitle

\section{Introduction}

The ATLAS and CMS Collaborations at the LHC recently reported ~\citep{ATLAS:2023fsd,CMS:2024pts} observations of entanglement between the spin degrees of freedom of top and anti-top ($t$ and $\bar t$) quarks. The experiments found a level of spin correlations in the $t \bar t$ pair that exceeded the Peres-Horodecki bound \cite{Peres1996, Horodecki1996,Afik:2020onf} which all non-entangled systems must respect. These results prove that the foundations of Quantum Mechanics (QM) can be successfully investigated at the TeV scale, and also in systems with unstable particles. 

In addition to observing entanglement, LHC experiments will also seek to observe Bell Inequality violations (BIV)~\cite{Bell1964,CHSH1969,BrunerRMP2014}, in $t \bar t$ pairs and other high-energy two-particle systems \cite{Fabbrichesi:2021npl, Severi:2021cnj, Han:2023fci, Fabbri:2023ncz, Barr:2021zcp, Fabbrichesi:2023cev, Bi:2023uop}.
BIVs provide proof of the non-classicality of quantum theory, by ruling out its compatibility with local hidden variable theories. 
 Since not all entangled states violate a Bell inequality, BIVs are considered a stronger form of correlations. 

A number of loopholes has been identified in connection with an experimental demonstration of BIVs~\cite{Aspect1982,Weihs1998,Scheidl2010}. In this work we investigate the possibility of closing the so-called ``locality loophole''~\cite{Aspect1975,Aspect1976}. It has been proven that a system of two parties,  obeying the laws of classical physics, can exactly reproduce the outcome of a quantum measurement on a maximally entangled state with the exchange of just a few bits of classical information~ \cite{Brassard:1999kj}. 
Therefore, we should ask ourselves what constitutes a convincing observation of non-classical properties in a collider setting, where the absence of communication is not guaranteed. 

Ideally, experimenters should restrict their study to events, where the QM measurements are space-like separated, to enforce that no classical information can be exchanged between the particles under consideration. 
However, in a collider setting this cannot be accomplished on an event-by-event basis.
In this work we argue for criteria to enforce such condition on average in an ensemble of $t \bar t$ events considered for a particular analysis. Of course, we have no reason to believe that top quarks, or any other pair of fundamental particles, actually possess the ability to exchange information during their lifetime, nor the ability to conspire with their partner to fake the presence of entanglement or Bell violations, when none is actually present; our argument simply attempts to close a loophole that is present in principle.
\begin{figure}
\vspace{-0.5 cm}
\includegraphics[width=0.48\textwidth]{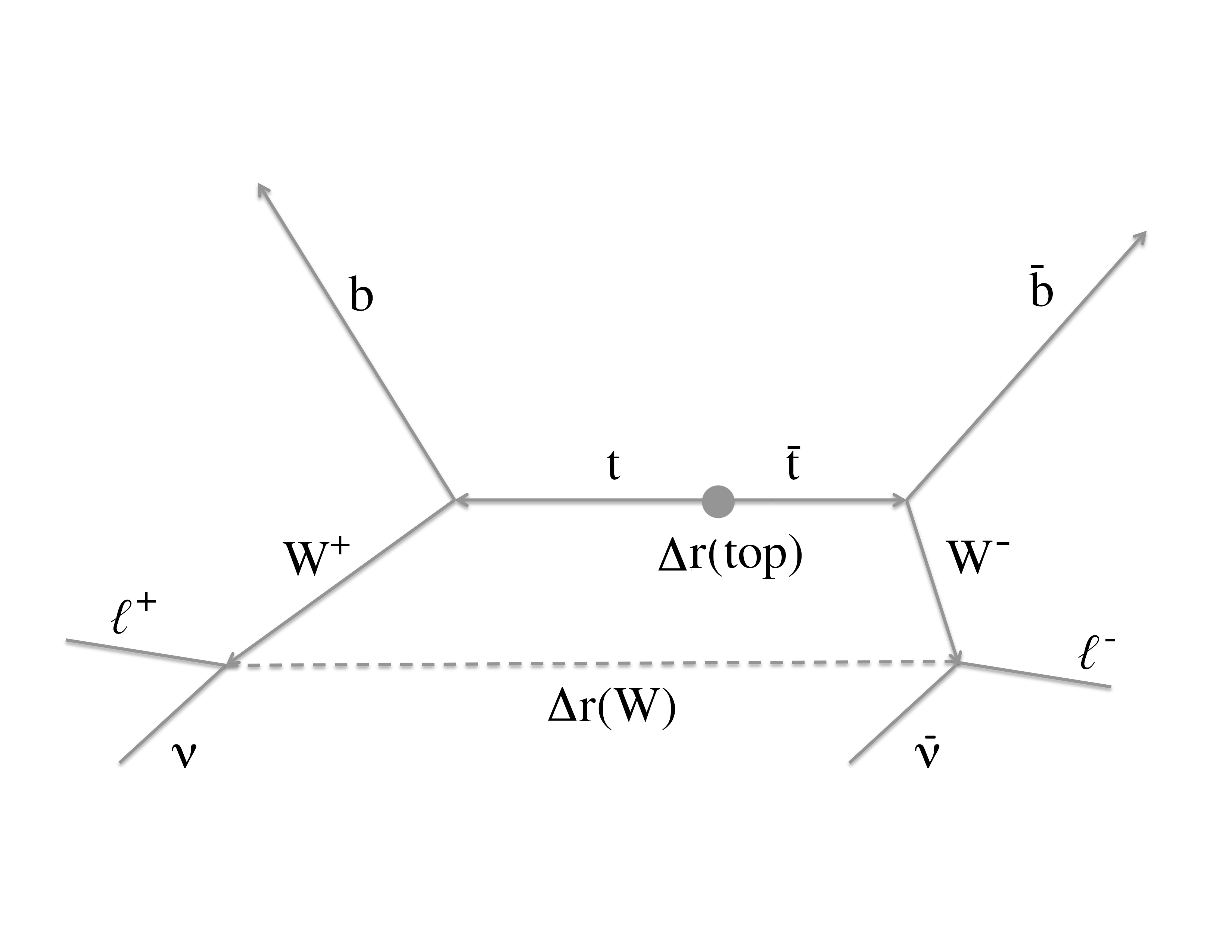}
\vspace{-1.2 cm}
\caption{ Schematics of $t\bar{t}$ production with subsequent decay to $W$ bosons and $b$ quarks. Each $W$ boson then decays to a charged lepton and a neutrino. The distance between the tops' decay is denoted $\Delta r(\text{top})$, while the distance between $W$'s decay is denoted $\Delta r(\text{W})$. }
\label{fig:diagram}
\end{figure}

The central question in this paradigm is: what constitutes a measurement? The schematics of the $t\bar{t}$ production and decay is shown in Fig~\ref{fig:diagram}. Which interaction should be considered as the act of measurement? Spin correlations, and therefore entanglement and BIV, are measured via angular correlations in the top quarks' decay products \citep{Severi:2021cnj,Dong:2023xiw,Barr:2024djo}. Are these angles determined in the moment top quarks decay? Is it, instead, the moment $W^\pm$ bosons decay that constitutes the quantum measurement, since the particles that are actually observed, leptons and light jets, are produced at that stage? Is it the moment of the first contact of these final decay products with the macroscopic apparatus~\citep{Afik:2024uif}? The argument can continue, from the moment, when the pixel detector integrates the charge of the first hit, to the moment, when the first person looks at a plot. Without getting into a philosophical argument, here we will derive criteria based on 
the requirements of space-like separation of the following moments: the decays of the top and antitop quarks, the decays of $W$ bosons, and the first contacts of leptons with the macroscopic apparatus. There is also a possibility that space-like separated top quark decays produce $W$ bosons that become time-like separate at the moment of their decay.  To exclude this possibility we also consider a requirement that both types of decays are space-like separated. Out of these criteria we will recommend the most stringent one. 

A couple of remarks on the assumptions made in the presented analysis are prudent. First, we treat short-lived top quarks and $W$ bosons as particles, not as fields. Second, we assume that the location of the decay can be inferred from the kinematics of the particle, while strictly speaking, the Heisenberg uncertainty principle should be applied. It should be noted though that both counter arguments are based on a quantum mechanical treatment of the system, while we are trying to refute the classical explanation of entanglement. Third, the fact that the particles are space-like separated at the decay point does not exclude a possibility that their light cones overlapped in the past. This is always true for particles moving at subluminous speeds, which were in interaction at some point. Despite being in contact the system remains in a mixed state at least until one of the top quarks decays, hence the points of particles decays must be space-like separated to exclude causal contact.   
%
%

This work is organized as follows. After a brief introduction to top spin correlations and entanglement, in Section~\ref{Sec:top_spin} we propose criteria to close the locality loophole based on $f$, the fraction of events where the measurements on the top and anti-top quarks are space-like separated. Next, in Section~\ref{Sec:space_like_fraction} we evaluate $f$, in the four definitions described above. For the "top" and "W" definitions we use Monte Carlo simulation of the top pair production and decay kinematics. Since this is a straightforward task, we propose that analyzers use their own best MC simulations to determine the relevant space-like fractions. This way detector efficiency and resolution will also be taken into account. For the "top" and "lepton" definitions we will also show analytical calculations of $f$. We conclude in Section~\ref{Sec:Conclusion}.
\section{Top spin correlations, entanglement, and Bell violations} 
\label{Sec:top_spin}

Let us give a quick overview of the strategy used to measure spin correlations, and the corresponding conditions for observing entanglement and Bell inequality violations. The correlation between the spin orientations  of top and anti-top quarks is described by the spin correlation matrix $C_{ij}$, which directly enters the top/anti-top spin density matrix,
\begin{equation}
    \rho = \frac{ \mathbf{1} \otimes \mathbf{1} + B_{1i} \,  \sigma_i \otimes \mathbf{1} + B_{2j} \,  \mathbf{1} \otimes \sigma_j  + C_{ij} \, \sigma_i \otimes \sigma_j }{4}, \label{rho}
\end{equation}
where $i,j = 1,2,3$ and $\sigma$ are the Pauli matrices. The vectors $B_{1i}$ and $B_{2j}$ represent the spin polarization of the top and anti-top quarks, while the $C_{ij}$ matrix parameterizes their spin correlations. The angular distribution of the top and anti-top quark decay products is used to evaluate $C_{ij}$ experimentally, and therefore extract spin correlations. In some regions of phase space, correlations are so strong that they can only be explained by entanglement.  According to the Peres-Horodecki criterion the system is entangled if~\cite{Bernreuther:2013aga}
\begin{equation}
\Delta_E=C_{nn}+|C_{rr}+C_{kk}|>1, \  
\label{eq:Peres_Horodecki}
\end{equation}
where $C_{nn}, C_{rr}$ and $C_{kk}$ are the diagonals of $C_{ij}$ in the helicity basis $\lbrace k, r, n \rbrace$. An even stronger quantum connection is manifested in the violation of Bell inequalities. It is convenient to consider the CHSH formulation, which in our case reduces to \cite{Severi:2021cnj}:
\begin{equation}
B_\pm = |{C_{rr} \pm C_{nn}|>\sqrt{2}} \ , 
\label{eq:CHSH}
\end{equation}
where either condition is sufficient to establish a violation.
Let's assume we know how to calculate $f$, the fraction of $t \bar t$ events where the quantum measurements of spin take place at space-like separation. In \cite{CMS-PAS-TOP-23-007}, a proposal was made to extend the condition in Eq.~~\ref{eq:Peres_Horodecki} to ensure its value cannot be explained via classical communication. Here we continue the same argument to $B_\pm$, the Bell markers of Eq.~\ref{eq:CHSH}, proceeding as follows: the time-like separated events are assumed to have maximum correlations, and therefore the maximum mathematically allowed values of $\Delta_E^{\text{max}}=3$ and $B_\pm^{\text{max}}=2$; the space-like separated events are assumed to have the maximum allowed "classical" values, {\it i.e.}~those that do not require entanglement or BIV, $\Delta_E^{\text{class}}=1$ and $B_\pm^{\text{class}}=\sqrt{2}$. Then, with $f$ being the fraction of events where the measuremens are space-like separated, the largest values that could be explained by classical communication are: 
\begin{equation}
\Delta_E^{\star}=f \Delta_E^{\text{class}} +(1-f) \Delta_E^{\text{max}}=3-2f\ 
\label{eq:DEcritical}
\end{equation}
for entanglement, and 
 \begin{equation}
B_\pm^{\star}=f B_\pm^{\text{class}} +(1-f) B_\pm^{\text{max}}=\sqrt{2}f+2(1-f)\ 
\label{eq:BEcritical}
\end{equation}
for Bell inequality violation. 
\section{Space-like fraction} 
\label{Sec:space_like_fraction}
All that is left, at this point, is evaluating the probability $f$ that an event is space-like. Let $\tau_1 = (ct_1, x_1, y_1, z_1)$ and $\tau_2 = (ct_2, x_2, y_2, z_2)$ be the four dimensional coordinates of the measurements of the top/anti-top quark spin orientation. To guarantee that the connection between the top and anti-top quarks is of quantum nature, we need to make sure that these points are separated by a space-like interval 
\begin{equation}
(x_1-x_2)^2+ (y_1-y_2)^2+(z_1-z_2)^2- (ct_1-ct_2)^2>0 \ . 
\label{eq:interval}
\end{equation}
The values of $\tau_1$ and $\tau_2$ depend on the chosen definition of quantum measurement, and we will investigate each one in detail next. It is important to remember that experimentally we do not measure the top quark or the $W$ boson decay length. Thus, in the cases of top quark and $W$ boson decays, we can only determine $f$ on statistical basis. This is not an issue, however, since the entanglement and BIV markers are also determined as averages on an ensemble of events. In the lepton contact definition the directions of leptons and the location of the their production vertex are known event-by-event, hence the interval between first contacts with the beam pipe can in principle be extracted for each event. (We will show that in practice this is not necessary.) 

\subsection{The top quark decay}
\begin{figure*}
\begin{center}
\includegraphics[width=0.32\textwidth]{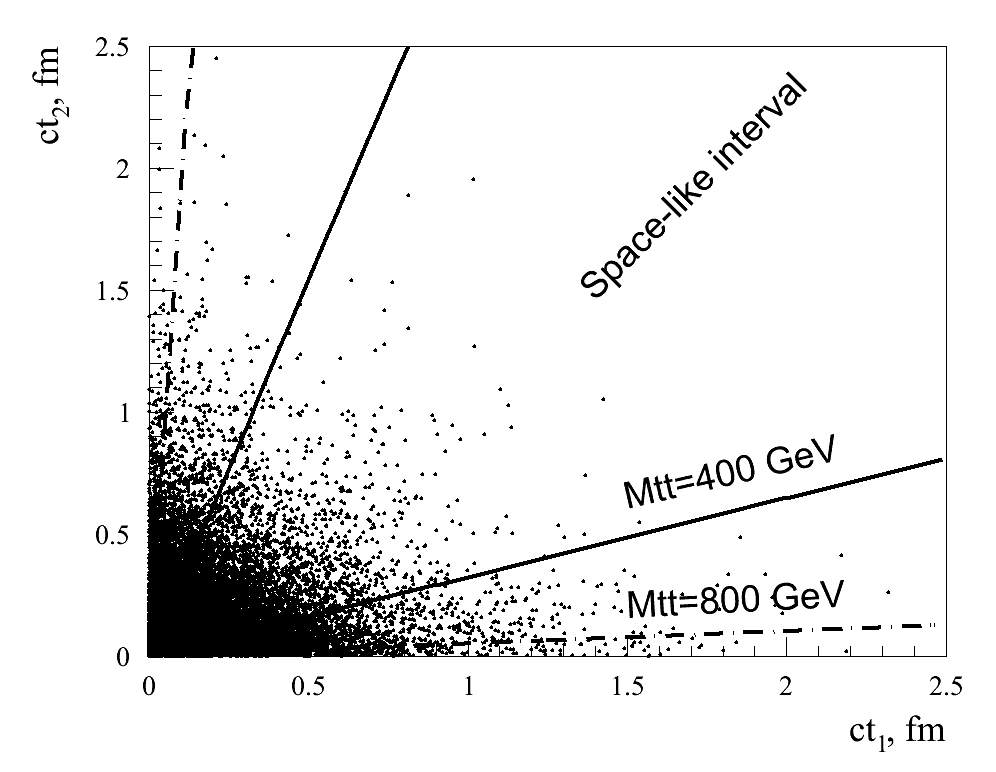}
\includegraphics[width=0.32\textwidth]{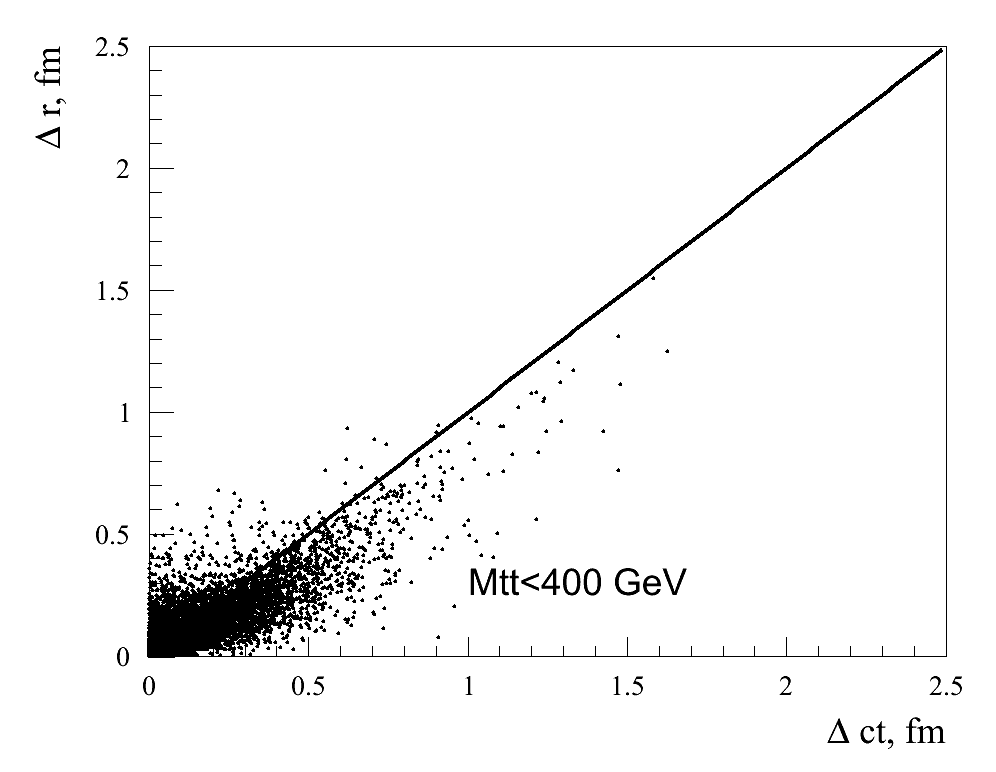}
\includegraphics[width=0.32\textwidth]{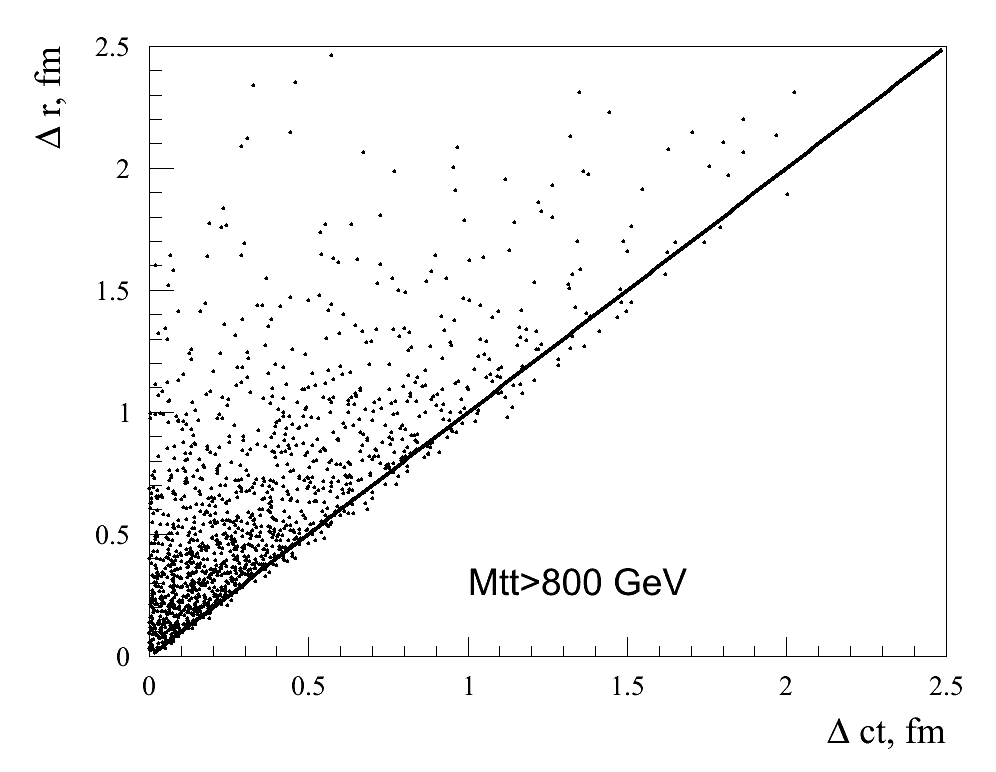}
\includegraphics[width=0.32\textwidth]{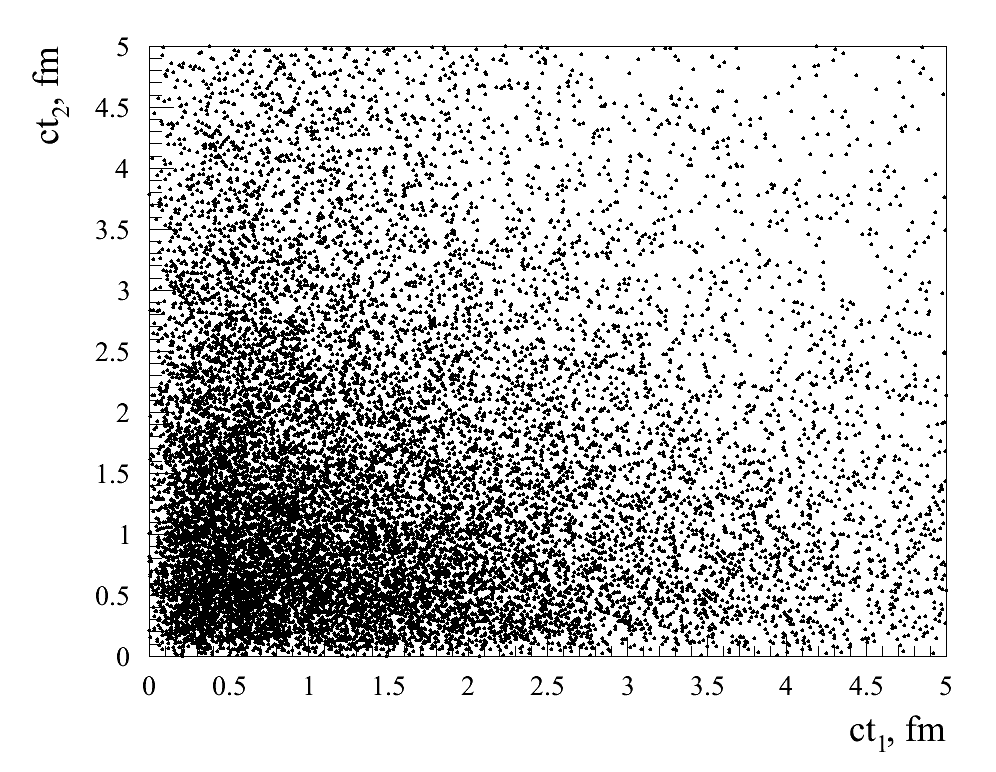}
\includegraphics[width=0.32\textwidth]{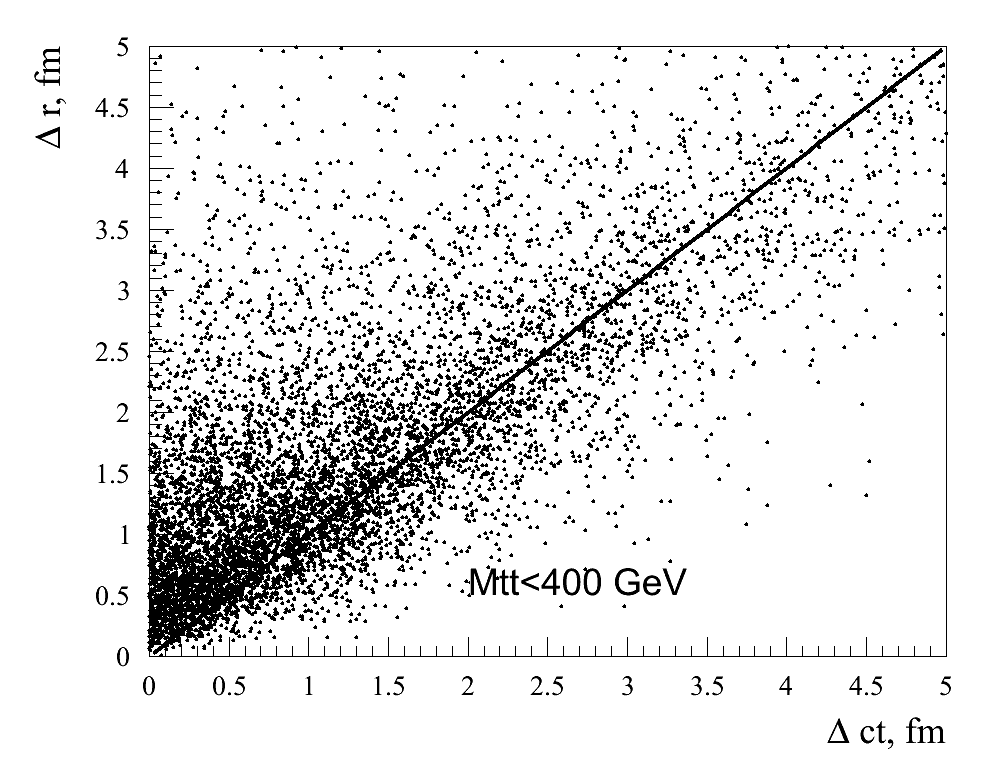}
\includegraphics[width=0.32\textwidth]{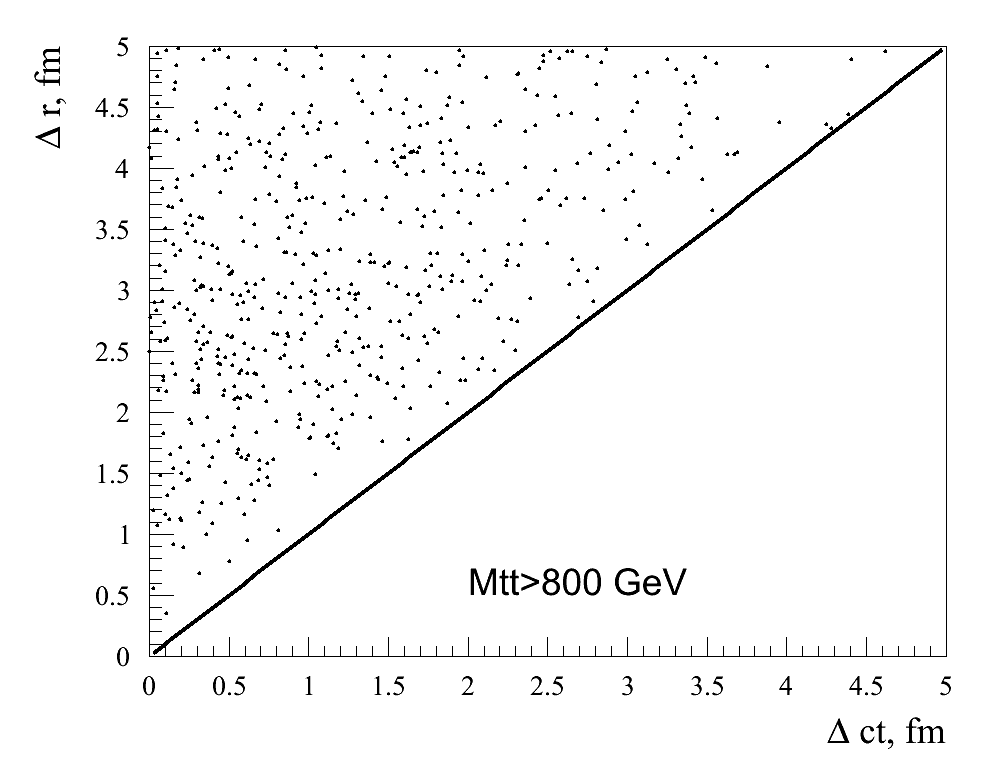}
\caption{Top row: top quark decays, bottom row: $W$ boson decays.  Left: time from production to  decay of particle vs that of anti-particle.  Lines show the limits of space-like interval for $M_{t\bar{t}}=400$ GeV (solid) and $M_{t\bar{t}}=800$ GeV (dashed). Center and right: distance vs time difference between particle decays for $M_{t\bar{t}}<400$ GeV (center) and $M_{t\bar{t}}>800$ GeV (right). Events below the diagonal are time-like separated, while the ones above the diagonal are space-like separated.}
\label{fig:t1t2_r_ct}
\end{center}
\end{figure*}
Let us consider the center of mass of the top and anti-top quark system. In this reference frame the velocities of the two particles are equal in magnitude and opposite in direction $\vec{\beta_t}=-\vec{\beta}_{\bar{t}}$. For brevity we shall refer to the magnitude of these vectors as $\beta$, which is related to the invariant mass of the $t\bar{t}$ system, $M_{t\bar{t}}$, through the simple relation 
\begin{equation}
\beta=\sqrt{1-(2m_t/M_{t\bar{t}})^2} \ , 
\label{eq:beta_vs+Mtt}
\end{equation}
where $m_t$ is the top quark mass.  The corresponding relativistic factor is $\gamma=1/\sqrt{1-\beta^2}$.

The top quark is an unstable particle, and its lifetime is exponentially distributed with a time constant corresponding to the top width $\Gamma_t=1.42$ GeV. We can direct the $z$ axis along the direction of flight of the top quark, so that the 4D coordinates of the top/anti-top decay points are $(\gamma ct_1,0,0, \beta \gamma ct_1)$ and $(\gamma ct_2,0,0, -\beta \gamma ct_2)$. To guarantee that the connection between the top and anti-top quarks is of quantum nature we need to make sure that their decays are separated by a space-like interval,
\begin{equation}
(\beta \gamma ct_1+\beta  \gamma ct_2)^2- (\gamma ct_1-\gamma ct_2)^2>0 \ . 
\label{eq:int_top}
\end{equation}
Thus, the condition for space-like separation is 
\begin{equation}
 \frac{1-\beta}{1+\beta} t_1 <t_2< \frac{1+\beta}{1-\beta} t_1\ . 
\label{eq:t1t2}
\end{equation}
This is shown by the solid lines corresponding to different values of $M_{t\bar{t}}$ in Fig.~\ref{fig:t1t2_r_ct} (top left), obtained via a Monte Carlo simulation accurate at NLO in QCD using POWHEG~\citep{Campbell:2014kua}. The events are distributed according to the survival probability of each top that decays exponentially but is boosted by the time dilation factor $\gamma$. Even though it is impossible to say on an event-by-event basis when the decays are time-like or space-like separated, we can approach the problem statistically. Integrating the probability density function over the area between the solid lines we find a simple relation, 
\begin{equation}
 f=\beta \ . 
\label{eq:Fspace}
\end{equation}
Note, that this fraction does not depend on the actual decay times. Since $\beta$ depends on $M_{t\bar{t}}$ so does $f$, and indeed for $M_{t\bar{t}}$ going to infinity we find that $f$ approaches 1. 
To appreciate the dependence of $\beta$ on $M_{t \bar t}$, in Fig.~\ref{fig:t1t2_r_ct} we also split $t \bar t$ phase space in the regions $M_{t\bar{t}} < 400$ GeV (top center) and $M_{t\bar{t}} > 800$ GeV (top right). The fraction of the space-like separated events, above the diagonal, is significantly lower for the events near production threshold compared to those at high invariant mass. 
\subsection{The $W$ boson decay}
Next we consider the definition of the quantum measurement based on the $W$ boson decays. The lifetime of the $W$ boson is exponentially distributed with a time constant 
corresponding to its width, $\Gamma_W=2.085$ GeV. The two measurements are then separated by the time and distance of the top quark decays plus that of the $W$ boson decays, as shown schematically in Fig.~\ref{fig:diagram}. The time elapsed from the $t \bar t$ production to decays of the $W^{+}$ and $W^{-}$ bosons, also obtained via a Monte Carlo simulation, is shown in Fig.~\ref{fig:t1t2_r_ct} (bottom left). The distance separation vs time separation between $W$ boson decays is also shown for events with $M_{t\bar{t}}<400$ GeV (bottom center) and $M_{t\bar{t}}>800$ GeV (bottom right). Similar to the top quark decays we observe that the probability of space-like separation is significantly lower for events near $t \bar t$ production threshold compared to those at high invariant mass. 

It is important to note that in this case more events are above the diagonal, that is, space-like separated, compared to what we had found for the top quark decay definition. We will come back to this point.

\subsection{Leptons hitting the beam pipe}
While top quark and $W$ boson decays happen on the scale of a femtometer, $1$ fm $=10^{-15}$ m, contact of leptons (or jets produced by light quarks) with the macroscopic apparatus, presumably the beam pipe, happens on the scale of 1 cm, so leptons from $t \bar t$ decay can be considered originating from the same point. Since leptons (and hadrons) have negligible mass compared to the LHC collision energy, they can be considered as moving at the speed of light. Let $l_1$ / $l_2$ be the distance from the production vertex to the positions of first contact of the leptons with the beam pipe, and $t_1$ / $t_2$ be the corresponding times. 
Let $\chi$ be the angle between the direction of flight of the two particles. Then the distance between their first contacts is 
\begin{equation}
 \Delta l= \sqrt{l_1^2+l_2^2-2l_1l_2 \cos \chi} \ , 
\label{eq:Deltal}
\end{equation}
and the time between the two events is 
\begin{equation}
 \Delta t= |l_1/c-l_2/c |\ . 
\label{eq:Deltat}
\end{equation}
Thus, the interval between the two events is
\begin{equation}
\Delta l^2- (c\Delta t)^2=2 l_1l_2(1-\cos \chi). 
\label{eq:interval_leptons}
\end{equation}
This quantity is clearly non-negative. 
Hence, if we take the perspective that the quantum measurements take place when leptons or jets contact the macroscopic apparatus around the interaction region, the locality loophole is automatically closed. This argument is similar to the one illustrated in Fig. 3 of~\citep{Ehataht:2023zzt} except the speed of electron or muon from top quark decay is much closer to the speed of light than the speed of $\tau$ lepton produced at $\sqrt{s}=10$ GeV, making the fraction of time-like events negligible.
\begin{figure*}
\begin{center}
\includegraphics[width=0.32\textwidth]{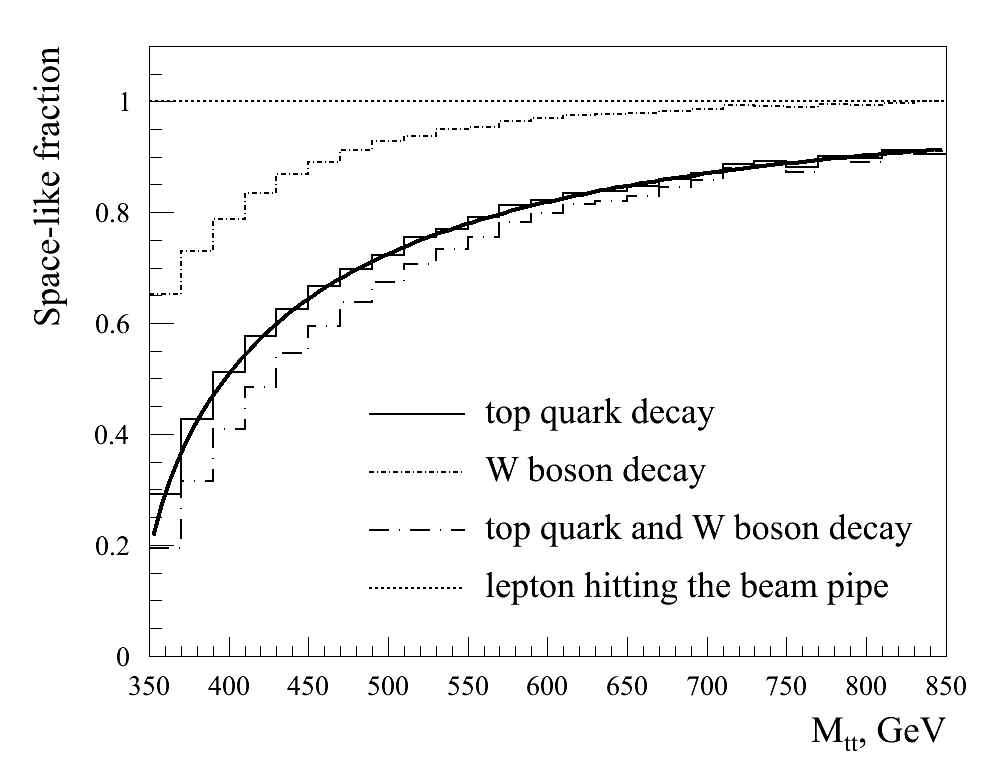}
\includegraphics[width=0.32\textwidth]{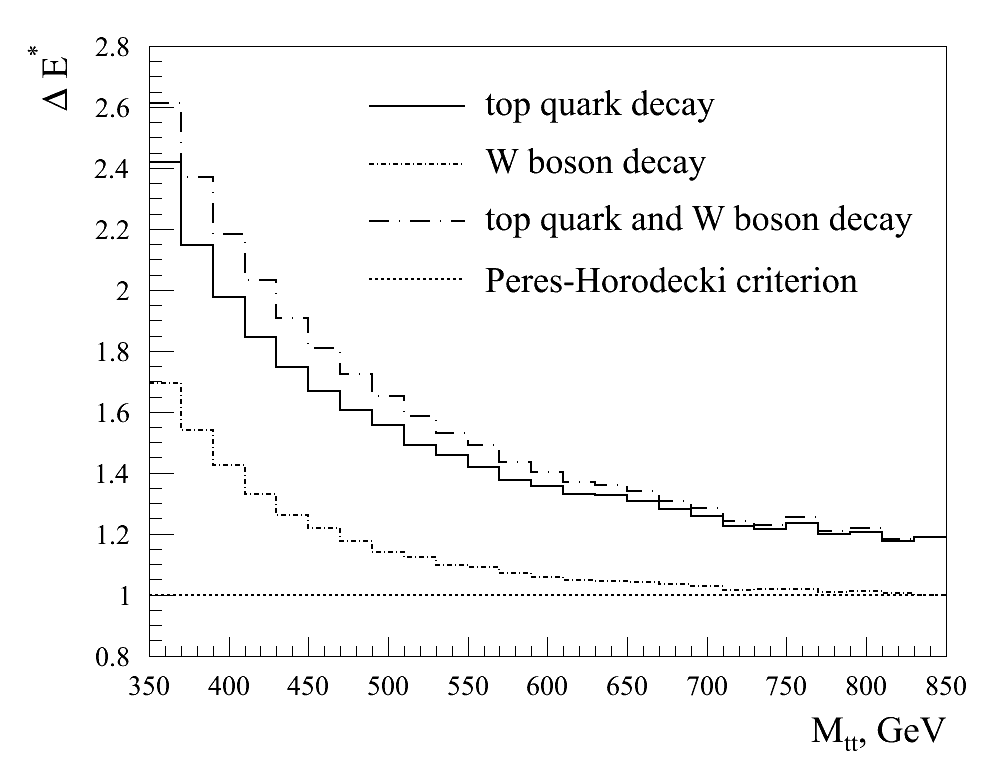}
\includegraphics[width=0.32\textwidth]{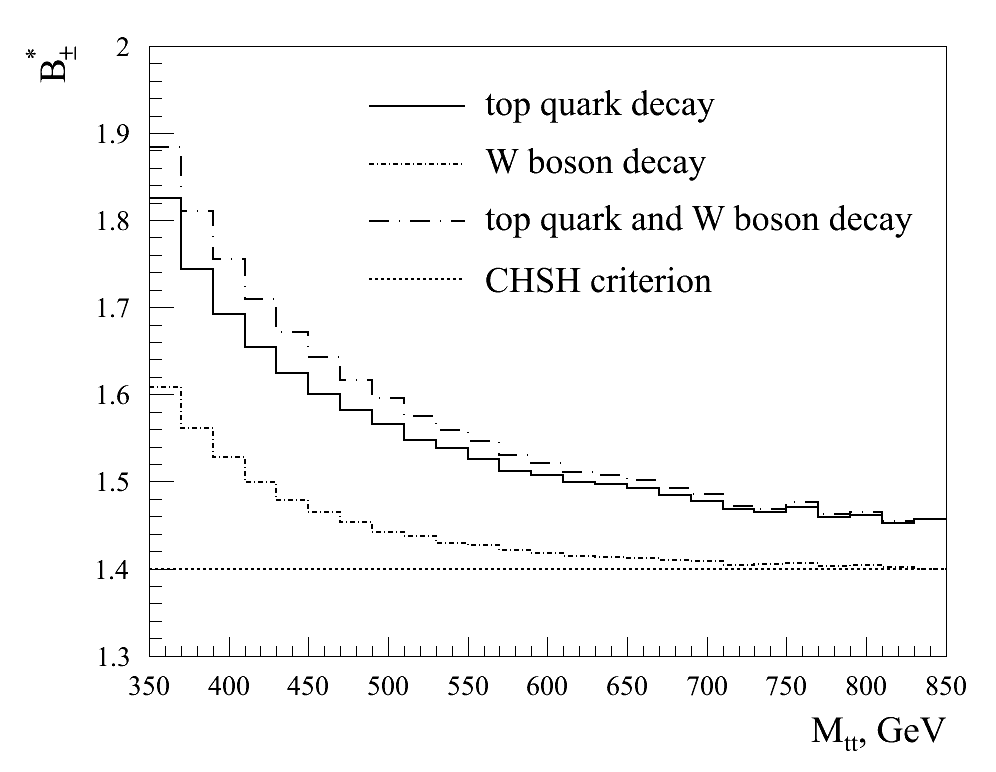}
\caption{ Legend inline. Left: Fraction of events with space-like separated measurements using four different definitions. The solid line shows the result of the analytical calculation, $f = \beta$, for the top quark decays. Center and right: Critical values of $\Delta_E$ (center) and $B_\pm$ (right) as a function of $M_{t\bar{t}}$ based on the considered definitions. The dotted  lines show the critical values in the absence of our proposed corrections and coincide with the definition based on lepton contact. }
\label{fig:Condition}
\end{center}
\end{figure*}
\subsection{Summary}
 In Fig.~\ref{fig:Condition} (left)  we show the space-like fraction $f$ as a function of $M_{t\bar{t}}$ based on the top quark and $W$ boson decays and leptons hitting the beam pipe. We also include a more stringent requirement that both types of decays happen in a space-like interval. For low $M_{t\bar{t}}$ this requirement reduces the fraction of space-like events, while for $M_{t\bar{t}}>800$ GeV it essentially coincides with the requirement based on top quark decay. For comparison we also show the result of the analytical calculation for top quark decays, which is in excellent agreement with the simulation.   

Of course, we advocate for selecting the most stringent requirement of top quarks and $W$ bosons both being in the space-like interval, when evaluating $f$. Therefore, the critical values for the entanglement marker $\Delta_E$ and for the Bell markers $B_\pm$ become a function of $M_{t\bar{t}}$, and are shown in Fig.~\ref{fig:Condition} (center and right). If the observed value of $\Delta_E$ or $B_\pm$ exceeds these critical values with enough significance, we can say that the detection of entanglement or of Bell violations cannot be argued away by assuming classical communication. 
\section{Conclusion} 
\label{Sec:Conclusion}
In light of the observation of entanglement in $t\bar{t}$ pairs, of concrete prospects for detecting Bell violations, and in general for a future Quantum Information program at the LHC, it is crucial to carefully analyze our strategies, to ensure that such counter-intuitive physical phenomena are detected experimentally in a convincing manner. Part of this endeavour consists of making sure "alternative" explanations for quantum phenomena, however unlikely, are ruled out. We analyzed the locality loophole, and proposed criteria to establish that the observed values of $\Delta_E$ or $B_\pm$ cannot be explained by classical communication. We considered three different definitions for when the "quantum measurement" of spin takes place: at the top quark decay, at the $W$ boson decay, and at the lepton/jet contact with the macroscopic apparatus. We showed that the space-like fraction of events is the smallest, when requiring that both top quarks and $W$ bosons decay within space-like interval. For high invariant masses, typically required for the Bell violation this is almost identical to just the top quark decay requirement. We provided numerical values for our proposed criteria. We also show that if the measurement happens when particles hit the detector there is no locality issue at all. 
\section{Post Scriptum} 
\label{Sec:PS}
In the argument presented above one critical assumption is that the top quark decay time is not correlated with that of the antitop quark. Should that not be the case, {\it e.g.}~the events are clustered in the upper and lower wedges in Fig.~\ref{fig:t1t2_r_ct} (top left), while still individually following the exponential distributions, the locality loophole cannot be closed. Since measuring top quark decay length is out of the question, a system consisting of a pair of $B$-mesons or $\tau$ leptons  might be better suited for addressing this potential issue. 
\section*{Acknowledgements}
The authors thank Marcel Vos, Claudio Severi, Dorival Goncalves and Alan Barr for stimulating discussions. RD acknowledges support from the U.S. Department of Energy under the grant DE-SC0008475.  

\bibliography{biblio}

\end{document}